\documentclass[prl,letterpaper,twocolumn,showpacs,superscriptaddress]{revtex4}
\usepackage{graphicx,psfrag,amsmath,amssymb,amsfonts,bbm,latexsym,color,dcolumn}

\begin{document}
\title{Detectability of dissipative motion in quantum vacuum
via superradiance}

\author{Woo-Joong Kim}
\affiliation{Department of Physics and Astronomy,Dartmouth
College,6127 Wilder Laboratory,Hanover,NH 03755,USA}

\author{James Hayden Brownell}
\affiliation{Department of Physics and Astronomy,Dartmouth
College,6127 Wilder Laboratory,Hanover,NH 03755,USA}

\author{Roberto Onofrio}
\affiliation{Department of Physics and Astronomy,Dartmouth
College,6127 Wilder Laboratory,Hanover,NH 03755,USA}
\affiliation{Dipartimento di Fisica ``G. Galilei'',Universit\`a di
Padova,Via Marzolo 8,Padova 35131,Italy}
\date{\today}

\begin{abstract}
We propose an experiment for generating and detecting vacuum-induced
dissipative motion. A high frequency mechanical resonator driven in 
resonance is expected to dissipate mechanical energy in quantum vacuum 
via photon emission. The photons are stored in a high quality 
electromagnetic cavity and detected through their interaction with 
ultracold alkali-metal atoms prepared in an inverted population of hyperfine states.
Superradiant amplification of the generated photons results in 
a detectable radio-frequency signal temporally distinguishable from 
the expected background.
\end{abstract}
 
\maketitle
{\it Introduction} -
Macroscopic quantum effects are suitable to bridge the gap between
quantum theory and general relativity. In this context, observable 
effects due to the change in the boundary conditions of quantum fields, 
like the creation of particles in an expanding universe 
\cite{Schroedinger} or the Casimir force \cite{Casimir}, may 
provide crucial information.
So far the attention has been mainly focused on {\sl conservative}
Casimir forces, with measurements performed in a variety of geometries
ranging from the original parallel plane \cite{Sparnaay,Bressi} to the
sphere-plane \cite{vanBlokland,Lamoreaux,Mohideen,Chan,Decca} and
crossed-cylinders \cite{Ederth}. Meanwhile, there have been
several theoretical attempts at studying the {\sl dissipative}
contribution of vacuum fluctuations to understand their interplay
with the relativity of motion \cite{Moore,Fulling,Jaekel}.
In principle, dissipative Casimir forces should be evidenced as a further
damping source in the non-uniformly accelerated motion of micromechanical
resonators already implemented for measuring the conservative component of the force.
However, given the level of dissipations coming from more technical sources, a direct
detection of the dissipative Casimir force seems out of experimental reach.
Instead of focusing the attention on deviations from conservative motion, the
dissipation induced in vacuum could be more easily detected by looking at
the radiated photons that are less contaminated from other sources
of noise \cite{Lambrecht}. This phenomenon, also known as dynamical Casimir 
effect (see \cite{Barton} for an updated review), can be understood both as 
the creation of particles under nonadiabatic changes in the boundary conditions 
of quantum fields, or as classical parametric amplification where the zero 
point energy of a vacuum field mode is exponentially amplified in time.
Theoretical analysis indicates that under parametric amplification in an
electromagnetic cavity an initial state of $N_0$ photons with frequency
within the resonance bandwidth of the fundamental mode of the cavity $\omega$
is transformed into a squeezed state with an average number of
photons growing in time as \cite{Dodonov,Plunien,Crocce}
\begin{equation} \label{eq:photonnumber}
N_{\mathrm{Cas}}(t)=N_0 \sinh^{2}(\Omega \epsilon t),
\end{equation}
assuming that the parametric resonance condition with a mechanical driving at a 
frequency $\Omega=2 \omega$ is fulfilled.
The product $\Omega \epsilon$ represents the squeezing parameter, 
with the modulation depth $\epsilon=v/c$, where $v$ is the velocity of the resonator and $c$ 
the speed of light. This exponential growth is eventually limited by the photon leakage of the cavity
expressed through the optical quality  factor $Q_{\mathrm{opt}}$, which saturates
at the hold time $\tau=Q_{\mathrm{opt}}/\omega$, reaching a maximum photon population  
\begin{equation}\label{eq:sat}
N_{\mathrm{Cas}}^{\mathrm{max}}=N_{\mathrm{Cas}}(\tau)=N_0 \sinh^{2}(2 Q_{\mathrm{opt}} \epsilon).
\end{equation}

\begin{figure}[t]
\includegraphics[width=0.80\columnwidth,clip]{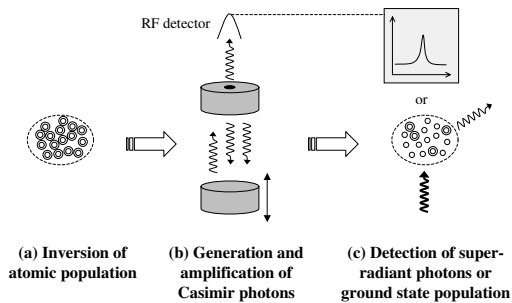}
\vspace{-0.5cm}
\caption{Generation and detection of photons irradiated through 
vacuum-induced damping of motion. The two-level
atoms are optically pumped to the maximum angular momentum state
of the spin-orbit manifold which only allows for a single,
downward, circularly polarized, magnetic dipole transition. The
atoms are then sent through the cavity (a). The
Casimir photons are generated through mechanical modulation of one
FBAR resonator (b). An amplified superradiant pulse is triggered
by a weak Casimir signal and detected by radio-frequency electronics 
or atomic fluorescence (c). } \label{fig1}
\end{figure}

In this Letter, we discuss a generation mechanism for
Casimir photons and a nearly quantum-limited
photodetection scheme in the radio-frequency range
based on the interaction of the generated photons with
an excited population of atoms.
This proposed experiment, initially sketched in \cite{Onofrio},
exploits in addition the high gain of superradiant emission to boost 
the expected signal to detectable levels, and a schematic outline
of its components is shown in Fig.~\ref{fig1}.
An extremely weak signal of Casimir photons will trigger the emission 
of an intense, time-compressed, superradiant pulse whose characteristic 
delay time will provide the signature of mechanically induced vacuum radiation.

{\it Generation of Casimir photons} - Current thin film technology makes feasible 
mechanical motions in the GHz range, with the highest frequency reported 
to date $\Omega/2\pi=$3.0 GHz and a modulation depth of 
$\epsilon=10^{-8}$ \cite{Zhang}.
This has been obtained through a film bulk acoustic resonator (FBAR)
\cite{Ruby,Cleland,Gabl}, consisting of a vibrating aluminum
nitride (AlN) film of thickness corresponding to one half of the
acoustic wavelength, sandwiched between two electrodes.
The average number of photons at saturation, Eq.~(\ref{eq:sat}),
depends on the product of two parameters, $Q_{\mathrm {opt}}$ and
$\epsilon$, which can be on the order of $10^8$ and $10^{-8}$
respectively. The average number of photons in the cavity is
very sensitive to this product, with
$N_{\mathrm{Cas}}^{\mathrm{max}} = 1.4, 13, 740$ for values of
$Q_{\mathrm {opt}} \epsilon = 0.5, 1, 2$ for a vacuum state 
with $N_0=1$, respectively.
The expected saturated power initiated by Casimir emission is
\begin{equation} \label{eq:phpower}
P_{\mathrm{Cas}}=
N_{\mathrm{Cas}}^{\mathrm{max}} \frac{\hbar \omega}{\tau}.
\end{equation}
For a 3.0 GHz FBAR resonator and a benchmark value of 
$Q_{\mathrm {opt}} \epsilon \simeq 1$ at the edge of current technology, the 
saturated power $3 \times 10^{-22}$ W is too low to be directly detectable. 
This demands the use of an efficient, nearly quantum-limited, photon detector in 
the radio-frequency range.

{\it Detection of Casimir photons} - Ultra-sensitive atomic
detection schemes can be exploited for detecting Casimir photons
by preparing an ensemble of population-inverted atoms in a hyperfine
state whose transition frequency corresponds to the cavity resonance.
An additional amplification process is available in which the weak Casimir 
signal triggers the stimulated emission of the ensemble of atoms.  
This effect is a form of superradiance \cite{Benedict,Gross}.
One favorable feature to be exploited for the proposed scheme is that 
the hyperfine splitting of the ground states for alkali atoms ranges 
from 0.2 GHz for Li to 9 GHz for Cs, conveniently matching the operating 
frequencies of FBAR resonators achieved or achievable in the 
near future \cite{Note1}.
The hyperfine transition in the ground state occurs through a
magnetic dipole interaction, and its natural lifetime in free
space is approximately
\begin{equation} \label{eq:T1}
T_1 \approx \frac{3 \pi \epsilon_0 \hbar c^5}{\mu_{B}^2 \omega^3},
\end{equation}
where $\mu_{B}$ is the Bohr magneton and $\epsilon_0$ the electric
permittivity in vacuum. This natural lifetime in
free space is favorably reduced inside a resonant cavity due to
the modification of density of states \cite{Purcell,Goy}
\begin{equation}
\label{eq:Tcav} T_1^\mathrm{cav} = \frac{4\pi^2}{3Q_\mathrm{opt}}
\frac{V}{\lambda^3} T_1,
\end{equation}
where $V$ is the cavity volume.  For a few GHz cavity with
1~cm$^2$ cross-sectional area and $Q_{\mathrm {opt}}=10^8$, the
natural lifetime is reduced by a factor of $10^{10}$.
In spite of this cavity-enhanced spontaneous rate, the typical
hyperfine transition lifetime for the alkali-metal atoms is still
impractically long, on the order of $10^3-10^{5}$~s.
The superradiant lifetime - the characteristic time scale for superradiant
evolution when $N_{\mathrm{at}}$ atoms are enclosed within the
cavity - is $T_{\mathrm{SR}} = T_1^{\mathrm{cav}}/N_{\mathrm{at}}$.
Hence the emission time scale for the experiment is further reduced
in the millisecond range for $N_{\mathrm {at}} \approx 10^8$ or less.
The peak power of the superradiant pulse is
\begin{equation} \label{eq:atpower}
P_{\mathrm {SR}} = N_{\mathrm{at}} \frac{\hbar \omega}{T_{\mathrm {SR}}},
\end {equation}
increasing quadratically with the number of atoms.
Considering as before a few GHz resonator with $10^8$ atoms and
$T_{\mathrm {SR}}=10^{-3}$ s, yields $P_{\mathrm{SR}} = 10^{-13}$~W,
a billionfold improvement over the initial power estimated in Eq.~\ref{eq:phpower}.

The superradiant emission can be detected in either of two ways.
First, a power or field detector can be coupled to the cavity. 
The detector should be fast enough to resolve one superradiant lifetime. 
Such direct measurement would be preferred although the coupling mechanism 
itself is likely to reduce the quality factor of the cavity significantly 
in order to attain sufficient coupling efficiency.
Micro-bolometers mounted on etched ``spiderwebs'' have an
ultimate sensitivity of
$10^{-16}\,\mathrm{W}/\sqrt{{\mathrm{Hz}}}$ in the GHz range
\cite{Turner}.  Spectrum analyzers are sensitive to sub-fW RF
power of kHz bandwidth \cite{Agilent}, and the temporal profile 
of the burst can be reconstructed through vector analysis.
Second, the exiting atoms can be interrogated resonantly with the
lower hyperfine state to ascertain the lower state population and
therefore the energy released into the cavity. Either D-line
fluorescence or ionization current can be monitored at the 
thousand atom level sensitivity \cite{Chen}. The average delay time 
can be inferred by varying the time the atoms stay within the cavity.
Coherent D-line excitation may also generate free-induction decay, due 
to the coherent magnetic moment developed on the hyperfine transition as 
a consequence of the amplification process, that would have a clearer signature.
The two detection techniques are complementary to each other, and could be 
used in coincidence to further reject spurious signals. 

{\it Background rejection} - 
Casimir-generated photons are not the only seed to trigger the 
stimulated amplification process. In particular, any atom decaying 
spontaneously will also trigger a superradiant burst. 
In this case, the process is more commonly known as superfluorescence.
The temporal envelope of the photon burst allows for discrimination
among the triggering sources.
Indeed, the average delay between the initial stimulation of the
atomic population depends on the number of atoms and resonant 
photons $N_{\mathrm{ph}}$ initially present \cite{Haroche}:
\begin {equation} \label{eq:delay}
T_{\mathrm {D}}=T_{\mathrm {SR}} \ {\ln} \biggl[
\frac{N_{\mathrm {at}}}{1+N_{\mathrm {ph}}}\biggr].
\end {equation}
The delay is typically around ten superradiant lifetimes but with
an inherent uncertainty due to quantum fluctuations.  Both the delay 
and its uncertainty shrink as the number of initial photons increases.
Measuring the delay can then indicate the number of initial
Casimir photons. 
Tailoring the atomic number can further distinguish the Casimir
signal from superfluorescent pulses.
In order for the superradiant pulse to develop fully, the growth
rate must exceed any decay process, which is primarily Doppler
dephasing in the atomic cloud, and the atoms must remain in the
interaction region for a time longer than the delay time. 
Then superfluorescence will be suppressed relative to Casimir
superradiance provided that the atoms will be removed from the
cavity after the expected Casimir delay time but prior to the
superfluorescence delay $T_{\mathrm {D}}^{(0)}$ [obtained 
with $N_{\mathrm{ph}}=0$ in Eq.~\ref{eq:delay}]. 

{\it Experimental approach} - 
The Casimir photon population is allowed to reach saturation before
introducing the prepared atoms. The atoms can be trapped and cooled with 
standard magneto-optical techniques, optically pumped and then transported
into the cavity via optical tweezers. The existing photons then trigger a coherent 
pulse so long as the superradiant delay time is less than the cavity hold time. 
The direct use of an atomic beam is prevented by the short interaction time available in this configuration.
While, based on Eq.~\ref{eq:atpower}, it looks advantageous to increase the number
of atoms, an upper bound is imposed by the necessity to resolve the delay time as in 
Eq.~7. A major advantage of the proposed scheme is that outside the
cavity the atoms are effectively inert due to the long hyperfine lifetime. 
Furthermore, the atoms are not resonant with the direct emission of 
photons at frequency $\Omega$ originating from antenna dipole irradiation due 
to the mechanical oscillation.
\begin{table}[t]
\begin{tabular}{|c|c|c|c|c|}
\hline \rule{0pt}{12pt}
 & $^6$Li & $^{23}$Na & $^{87}$Rb & $^{133}$Cs \\
\hline \hline \rule{0pt}{12pt}
$\nu$(GHz) & 0.228 & 1.77 & 6.83 & 9.19\\
\hline \rule{0pt}{12pt}
$L$(mm) & 657 & 84.6 & 21.9 & 16.3\\
\hline \rule{0pt}{12pt}
$T_1$(s) & $8.4 \times 10^{16}$ & $1.8 \times 10^{14}$ & $3.1 \times 10^{12}$ &$1.3 \times 10^{12}$\\
\hline \rule{0pt}{12pt}
$T_{1}^{\mathrm {cav}}$(s) & $3.2 \times 10^{5}$ & $4.1 \times 10^{4}$ &$1.1 \times 10^4$ & $8.0 \times 10^3$\\
\hline \rule{0pt}{12pt}
$N_{\mathrm{at}}^{\mathrm {max}}$ & $6.4 \times 10^8$ & $8.2 \times 10^7$ & $2.2 \times 10^7$ & $1.6 \times 10^7$\\
\hline \rule{0pt}{12pt}
$T_{\mathrm {D}}^{(0)}$(ms) & $10.1 $ & $9.1$ & $8.5$ & $8.3$\\
\hline \rule{0pt}{12pt}
$T_{\mathrm {D}}$ (ms) & $8.8$ &$7.8$ & $7.1$ &  $7.0$\\
\hline \rule{0pt}{12pt}
$P_{\mathrm {Cas}}$(W) & $2.8 \times 10^{-23}$ & $1.7 \times 10^{-21}$  & $2.5 \times 10^{-20}$ &$4.6 \times 10^{-20}$ \\
\hline \rule{0pt}{12pt}
$P_{\mathrm {SR}}$(W) & $\,1.9 \times 10^{-13}\,$ & $\,1.9 \times 10^{-13}\,$ & $\,2.0 \times 10^{-13}\,$ & $\,1.9 \times 10^{-13}\,$ \\
\hline
\end{tabular}
\caption{Summary of relevant parameters and time scales involved 
in the superradiance dynamics for different alkali-metal isotopes 
already cooled in the $\mu$K range. 
The maximum number of atoms $N_{\mathrm{at}}^{\mathrm {max}}$ is 
chosen so that $T_{\mathrm {SR}}$ equals the assumed detector 
response time of 0.5~ms. 
The length of the electromagnetic cavity is chosen to match the 
hyperfine transition $L=2 c/\nu$ at frequency $\nu=\omega/2\pi$.
Optical quality factor $Q_{\mathrm {opt}}=10^8$ of the cavity 
and modulation depth $\epsilon=10^{-8}$ are assumed, yielding 
$N_{\mathrm{Cas}}^{\mathrm{max}}=13$.}
\end{table}

The atom number, $N_{\mathrm{at}}$, and the interaction time are
the primary adjustable parameters.  The maximum sensitivity is 
obtained when the former is adjusted so that the superradiant
lifetime is comparable to the detector speed (or the transfer speed
for the interrogation technique) and the latter is slightly less
than the superfluorescent delay time.
In Table~1, we summarize the various time scales and photon
production rates involved in our proposed scheme for hyperfine 
transitions of different alkali-metal species.  
Lithium is not a practical candidate due to the large cavity size, 
whereas cesium and rubidium require mechanical frequencies 
not presently available.
In this regard, sodium looks promising instead, with many individual 
steps of our proposed experiment already demonstrated. 
Sodium atoms in the maximum amount of $10^8$ have been trapped at a 
temperature of $T=100\, n$K  in a Bose condensed state \cite{Na},
radio-frequency transitions between hyperfine states have been 
intentionally driven \cite{Gorlitz}, and superradiance phenomena 
have been observed \cite{Inouye}. 
Concerning the detection speed, both micro-bolometers and heterodyne 
receivers are fast enough to resolve the shortest achievable pulse.
The difference in delay times is a few $T_{\mathrm{SR}}$ and so 
to suppress the superfluorescence, the atom transfer time out of 
the cavity should be at most $T_{\mathrm{SR}}$.  
Given that the transfer rate with optical tweezers
is limited by the mechanical drive moving the focusing lens to
roughly 10 cm/s, corresponding to a transfer time of 100 ms in and 
out of the cavity \cite{Gustavson}, the number of atoms required is   
$N_{\mathrm{at}}=6 \times 10^5$, corresponding to a peak power of 
$2 \times 10^{-15}$ W.

The key parameter in our scheme is the optical quality factor.
Assuming a relative error in the determination of the delay 
time of 10$\%$, a situation with $Q_{\mathrm {opt}} \epsilon=1$ 
determines the borderline for the temporal discrimination 
between superradiance induced by Casimir photons and superfluorescence, 
with a significantly improved signal for $Q_{\mathrm {opt}} \epsilon$ 
larger than unity. Quality factors of $10^8$ \cite{Raimond} and 
$10^{10}$ \cite{Rempe} have been reported for open, Gaussian, 
superconducting cavities. While, as already mentioned, the transient 
decrease of the cavity transmission occurs on a timescale longer than 
the superradiant pulse, care must be taken to minimize the losses 
introduced by the FBAR resonator as well as ports to admit the 
atoms and monitor the RF power. 

Scaling the size of existing resonators and dissipation of heat
are other issues to be carefully addressed. Given a typical size
of current FBARs of $\simeq 500 \mu$m${}^2$, increasing the FBAR
to 1 cm${}^2$ could adversely introduce additional acoustic modes
as well as enhance the risk of a pinhole breakdown through the AlN
film. The power required to drive the FBAR is obtained
by considering the kinetic energy of a vibrating material whose
energy is dissipated in the timescale of $Q_{\mathrm{m}}/\Omega$, 
where $Q_{\mathrm{m}}$ is the mechanical quality
factor, which gives $P=\rho V \Omega^3 \delta x^2/4
Q_{\mathrm{m}}$. Expressing the volume of the vibrating body  {\it
V} in terms of the cross sectional area {\it A} and the thickness
of the material, one half of the acoustic wavelength $2 \pi v_a/ \Omega
$, we obtain $P_{\mathrm{FBAR}}=\rho A v_a \pi^3 \epsilon^2
c^2/Q_{\mathrm{m}}$, independent of frequency. For a cross
sectional area of 1 $\mathrm{cm}^2$, $\epsilon=10^{-9}$, $\rho=10^3~$ kg/m${}^3$ 
and $v_a=10,400$ m/s for aluminum nitride with a typical mechanical 
quality factor $Q_{\mathrm{m}}=10^3$, the dissipated power is 
about 3 W \cite{Note2}, smaller than the maximum threshold power 
of $\simeq$ 10 W  applicable to a FBAR resonator without damaging it 
\cite{Ballou}. In principle, the mechanical quality factor can be 
increased up to 4,000 at room temperature, even larger at cryogenic temperatures, by
a careful design of the multiple reflection layer and the
refinement of an annealing process \cite{Multilayer}.
To minimize the moving boundary area to reduce heat load
and fabrication difficulty, the cavity could be a hollow, coaxial
waveguide terminated with length equal to half the resonant wavelength.
Finally, the thermal contribution to the initial photon population
$N_0$ at 10~mK is $N_{\mathrm{therm}}=6 \times 10^{-4}$, negligible 
with respect to the Casimir contribution \cite{Lambrechthermal}.

{\it Conclusions} - We have proposed an experiment involving 
superradiant amplification to detect the dynamical Casimir 
photons generated by a vibrating wall in an electromagnetic cavity.
Although the observation of radiated photons is limited by the
current technology, the use of superradiant atoms should 
overcome the technical limits and make their unambiguous detection
possible. The technology currently available for mechanical
resonators, the analysis of various alkali atoms, and the
interplay of the timescales indicate that a detection scheme based
upon use of Sodium atoms should have realistic chances to detect
photons radiated by non-uniformly accelerating bodies in quantum vacuum.

\begin{acknowledgments}
We thank D. A. R. Dalvit, E. S. Kim,  P. W. Milonni, and G. Ruoso for useful discussions.
\end{acknowledgments}

\end{document}